\documentclass[
amsmath,amssymb,
aps,
pra,
]{revtex4-2} 

\usepackage{graphicx}
\usepackage{dcolumn}
\usepackage{bm}
\usepackage[hidelinks]{hyperref}
\usepackage{amsmath}
\usepackage{amssymb}
\usepackage{physics}
\usepackage{tikz}
    \usetikzlibrary{arrows.meta}
    \usetikzlibrary{positioning}
    \usetikzlibrary{external}
\usepackage[caption=false]{subfig}
\usepackage{multirow}
\usepackage{array}
\usepackage{hhline}
\usepackage{booktabs}
\usepackage{siunitx}
\usepackage{xcolor}

\usepackage{xprintlen}
\newcommand\thefontsize{The current font size is: \f@size pt}

\newcommand{\I}{\ensuremath{\mathrm{I}}}
\newcommand{\X}{\ensuremath{\mathrm{X}}}
\newcommand{\Y}{\ensuremath{\mathrm{Y}}}
\newcommand{\Z}{\ensuremath{\mathrm{Z}}}

\definecolor{colortq1}{RGB}{34, 221, 137}

\begin{document}

\preprint{APS/123-QED}

\title{Method for noise-induced regularization in quantum neural networks}

\author{Viacheslav Kuzmin}
\author{Wilfrid Somogyi}
\author{Ekaterina Pankovets}
\author{Alexey Melnikov}
\thanks{Corresponding author, e-mail: alexey@melnikov.info
\begin{center}
\fbox{
\begin{minipage}{1\textwidth}
Please check the published version, which includes all the latest additions and corrections: 
Adv. Quantum Technol. 8:e00603, 2025, DOI: \href{https://doi.org/10.1002/qute.202400603}{10.1002/qute.202400603}
\end{minipage}
}
\end{center}
}
\affiliation{Terra Quantum AG, 9000 St.~Gallen, Switzerland}


\begin{abstract}

In the current quantum computing paradigm, significant focus is placed on the reduction or mitigation of quantum decoherence. When designing new quantum processing units, the general objective is to reduce the amount of noise qubits are subject to, and in algorithm design, a large effort is underway to provide scalable error correction or mitigation techniques. Yet some previous work has indicated that certain classes of quantum algorithms, such as quantum machine learning, may, in fact, be intrinsically robust to or even benefit from the presence of a small amount of noise. Here, we demonstrate that noise levels in quantum hardware can be effectively tuned to enhance the ability of quantum neural networks to generalize data, acting akin to regularisation in classical neural networks. As an example, we consider two regression tasks, where, by tuning the noise level in the circuit, we demonstrated improvement of the validation mean squared error loss. Moreover, we demonstrate the method’s effectiveness by numerically simulating quantum neural network training on a realistic model of a noisy superconducting quantum computer.

\end{abstract}

\maketitle


\section{Introduction}
\label{sec:introduction}

Recently, significant progress has been made in developing quantum processing units (QPUs)~\citep{Bravyi_2022, Bruzewicz_2019, Chatterjee_2021}. However, the current generation of QPUs is still characterized by relatively short coherence times. Executing algorithms on these so-called noisy intermediate-scale quantum (NISQ) devices~\cite{Preskill_2018,kordzanganeh2023benchmarking} often requires significant error mitigation to combat the effects of decoherence and noise-induced errors~\cite{namiki1997decoherence,fedichkin2003measures,fedorov2004evaluation}.

Variational quantum algorithms (VQAs) are designed to operate within the constraints of NISQ devices.  However, VQAs cannot completely ignore the noise that arises on QPUs. The headway has been made in understanding the effects of specific types of noise, including those described by phase-damping, depolarizing, and amplitude-damping channels, as well as leakage error, shot noise, and coherent Gaussian noise~\citep{zen_sim_2021, sko_rob_2023, azs_nav_2023, din_eva_2022}. While noise typically poses challenges to VQAs, it can also be advantageous for specific quantum algorithms. For instance, it has been shown that stochastic noise aids in avoiding saddle points~\cite{liu2022stochastic} in variational circuit optimization, quantum noise protects quantum
classifiers against adversarial attacks~\cite{PhysRevResearch.3.023153}, and noise reduces the number of gates in reservoir computing~\cite{domingo2023taking}. In this work, we further investigate the potential benefits of quantum noise by developing a practical method for regularizing quantum machine learning models using the noise naturally present in quantum hardware.

In classical machine learning, exploiting noise is known to be a tool to prevent so-called overfitting. The issue of overfitting is a well-recognized challenge, occurring when a model achieves strong performance on the training dataset but struggles to generalize the unseen data. Overfitting occurs when a model captures noise or irrelevant patterns in the training data rather than the underlying true patterns that generalize across datasets. To mitigate this issue, the strategies of adding noise into the data, weights, or gradients have been proposed~\citep{neco.1995.7.1.108, neco.1996.8.3.643, neelakantan2015adding}.
 
The observed effectiveness of noise in preventing overfitting in classical machine learning suggests its potential utility in mitigating overfitting in quantum neural networks~\cite{schuld2021machine,qml_review_2023} (QNNs). Recently, it has been demonstrated that adding noise to initial data or to the weights of QNN can act as a regularisation similar to its classical counterpart. For instance, in~\cite{PhysRevA.106.052421}, the approach of adding noise to the quantum kernel method was explored. The proposed method operates as follows: initial data undergoes processing through a quantum circuit, after which the density matrix is measured, and the outputs are subsequently fed into a linear model. It appeared that adding noise at the quantum circuit stage helps to reduce overfitting. Notably, within this method, noise is incorporated during the data encoding process before training, which corresponds to the introduction of noise into the initial data. The idea of using noise as a tool to mitigate overfitting in QNNs was also investigated in~\cite{Nguyen2020}. The QNN model employed in this paper involves tuning the parameters of the Hamiltonians to guide the evolution toward a state that yields a target expectation value. In this paper, noise is added to the initial data, and the noise model is implemented as a Gaussian perturbation of the elements of the initial density matrix. Furthermore, the authors investigated the scenario of introducing noise into the weights of the model. In both instances, their model exhibited improved generalization performance in the presence of noise.

\begin{figure*}
    \centering
    \includegraphics[width=.8\linewidth]{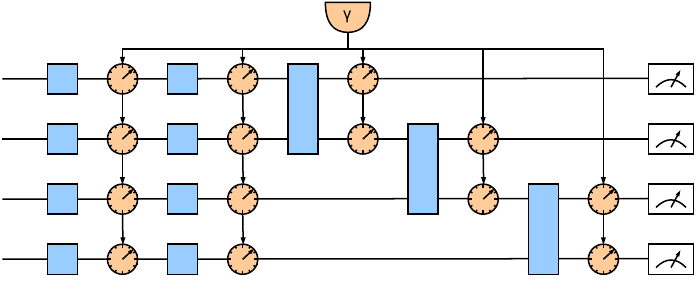}
    \caption{Scheme of the method for regularising quantum neural network models proposed in the work. Blue rectangles indicate single and two-qubit gates. Orange circles after each gate indicate the action of induced controllable noise with the strength optimized as a regularisation hyperparameter.}
    \label{fig:control_architecture}
\end{figure*}

In this paper, we complement the existing techniques by proposing an approach to leverage quantum noises, which naturally occur in quantum devices, to prevent overfitting. We suggest a method, schematically shown in Fig.~\ref{fig:control_architecture}, of adding one or several of the noise channels accessible in available quantum hardware. Adjusting the noise rates as hyper-parameters allows one to obtain a model that is superior to the noiseless scenario. We numerically investigate this method for a regression task by using several modeled noise channels and demonstrate improvement in the generalization performance. Finally, we discuss the practical implementation of the method in real quantum hardware and give evidence of its effectiveness by numerically simulating QNN training on a realistic noisy model of IBM's Kingston superconducting quantum computer.

\section{Decoherence}
\label{sec:decoherence}

In this section, the mathematical definitions are given for the three quantum noise channels studied in the paper. Namely, the amplitude-damping, phase-damping, and depolarizing channels. The relationship between these fundamental channels and the statistics commonly quoted to characterize quantum hardware are also described.

Decoherence arises as a result of the interaction between a quantum system and its environment. In the case of quantum processing units, the quantum state of the environment is not well known, and as a result, the state of the system qubits must be described in terms of the density matrix formalism. In general, the state of an $N$-qubit system interacting with its environment can be described by the density matrix $\rho$
\begin{equation}
    \rho = \sum_j p_j \op{\psi_j}{\psi_j}
\end{equation}
which denotes a probabilistic ensemble of pure states $\ket{\psi_j}$ where $p_j$ is the probability that the state of the system is $\ket{\psi_j}$. An evolution on this system can, in turn, be described in terms of a set of Kraus operators, $E_k$, which act on the density matrix as follows
\begin{equation}
    \mathcal{E}(\rho) = \sum_k E_k \rho E_k^\dagger
\end{equation}
where the set of operators $\{E_k\}$ describes the quantum operation $\mathcal{E}$, often referred to as a `quantum channel', and obeys the completeness relation $\sum_k E_k^\dagger E_k = \mathbb{I}$.

\textbf{Amplitude-damping (AD)} corresponds to \textit{energy loss} to the environment and describes the probability of the system in the state $\ket{1}$ decaying into the state $\ket{0}$. This channel is described by the following set of Kraus operators:

\begin{align}
    E_0 &=
    \begin{bmatrix}
        1 & 0 \\
        0 & \sqrt{1 - \gamma_\text{AD}}
    \end{bmatrix} \label{eq:ad_channel_e0}\\
    E_1 &=
    \begin{bmatrix}
        0 & \sqrt{\gamma_\text{AD}} \\
        0 & 0
    \end{bmatrix}
    \label{eq:ad_channel_e1}
\end{align}

\textbf{Phase-damping (PD)} represents a type of noise that destroys quantum coherence, and corresponds to a contraction of the Bloch sphere in the $x-y$ plane. This results in a destruction of quantum superpositions in the limit $\gamma_\text{PD} \rightarrow 1$, and a density matrix that describes entirely classical probability distributions of the states $\ket{0}$ and $\ket{1}$. The phase-damping channel is described by the following set of Kraus operators:

\begin{align}
    E_0 &=
    \begin{bmatrix}
        1 & 0 \\
        0 & \sqrt{1 - \gamma_\text{PD}}
    \end{bmatrix} \label{eq:pd_channel_e0}\\
    E_1 &=
    \begin{bmatrix}
        0 & 0 \\
        0 & \sqrt{\gamma_\text{PD}}
    \end{bmatrix} \label{eq:pd_channel_e1}
\end{align}

\textbf{Depolarizing (DP)} corresponds to the qubit experiencing an `error' with probability $\gamma_\text{DP}$. The possible types of error are the phase flip ($E_1$), bit flip ($E_2$), or phase-bit flip ($E_3$). In this case, the Kraus operators are the identity matrix and the three Pauli matrices:

\begin{align}
    E_0 &= \sqrt{1 - \gamma_\text{DP}}
    \begin{bmatrix}
        1 & 0 \\
        0 & 1
    \end{bmatrix} \label{eq:dp_channel_e0}\\
    E_1 &= \sqrt{\frac{\gamma_\text{DP}}{3}}
    \begin{bmatrix}
        1 & 0 \\
        0 & -1
    \end{bmatrix} \label{eq:dp_channel_e1}\\
    E_2 &= \sqrt{\frac{\gamma_\text{DP}}{3}}
    \begin{bmatrix}
        0 & 1 \\
        1 & 0
    \end{bmatrix} \label{eq:dp_channel_e2}\\
    E_3 &= \sqrt{\frac{\gamma_\text{DP}}{3}}
    \begin{bmatrix}
        0 & -i \\
        i & 0
    \end{bmatrix} \label{eq:dp_channel_e3}
\end{align}

\section{Training}
\label{sec:training}

\begin{figure*}
    \centering
    \includegraphics[]{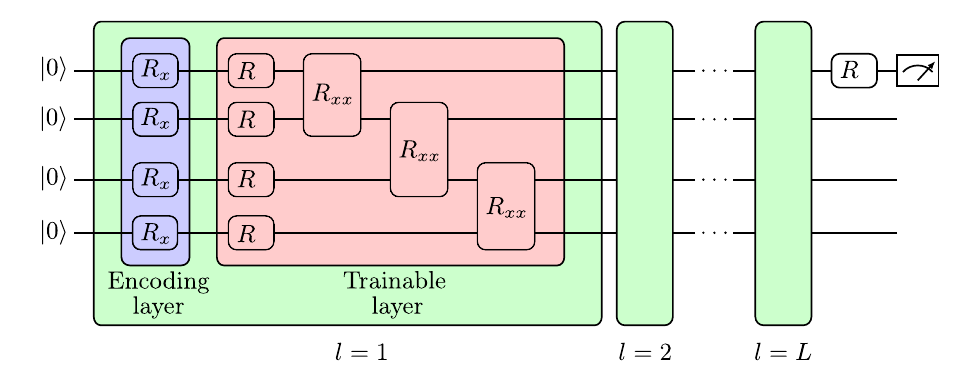}
    \caption{The ansatz circuit used to construct the QNN model, with alternating encoding and trainable layers, followed by a Pauli-Z measurement of the first qubit. Features are encoded with $R_x$ gates. The trainable layers are composed of a series of general rotation gates $R(\theta_1,\theta_2,\theta_3)$ followed by a layer of two-qubit $R_{xx}$ gates. The encoding and trainable layers are repeated $L$ times such that the input data is re-uploaded in each encoding layer. The measured qubit undergoes a parameterized general rotation before the measurement.}
    \label{fig:qnn_ansatz}
\end{figure*}

\begin{figure*}[p]
    \centering
    \includegraphics{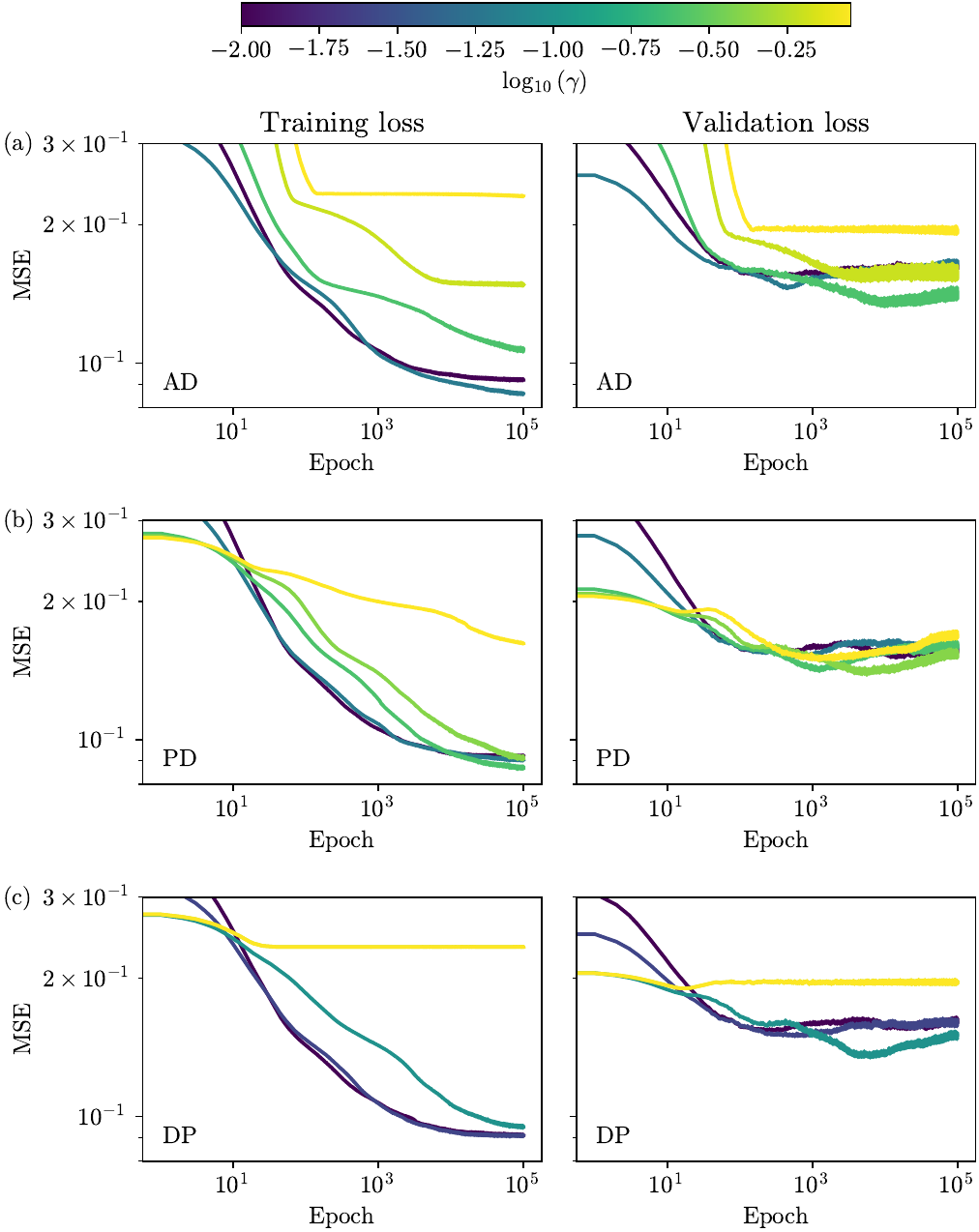}
    \caption{Mean MSE of the training (left) and validation (right) averaged over 20 training trajectories for the diabetes dataset.
    The plot shows results for QNN with various representative levels of $\gamma$ for each individual injected noise: (a) amplitude damping, (b) phase damping, and (c) depolarizing noise channels. The color of the lines indicates the noise level as given by the color bar. Trajectories with non-vanishing noise levels consistently achieve lower validation losses than those with vanishing noise.
    }
    \label{fig:epochal_loss}
\end{figure*}

We benchmark the noise-induced regularization technique using two regression benchmarks:

\begin{itemize}
  \item Diabetes (442 rows, 10 features).
Clinical baseline variables—age, body-mass index, mean arterial pressure, and seven serum-chemistry markers—are used to predict a quantitative disease-progression index measured one year later. The dataset ships with the \texttt{scikit-learn} Python package.
  \item Concrete Compressive Strength (1 030 rows, 8 features).
Each record describes a concrete mix by its ingredient masses (cement, blast-furnace slag, fly ash, water, super-plasticiser, coarse/fine aggregate) and curing age (days); the target is the resulting compressive strength in MPa.
\end{itemize}

Each dataset is randomly partitioned into $90\%$ training and $10\%$ validation samples.
A PCA model, fitted only on the training data, compresses the original predictors to four principal components. These components are linearly mapped to the interval $[0,2\pi]$, while the targets are scaled to $[-1,1]$. The PCA transform and scaling coefficients learned from the training split are then applied unchanged to the validation split, preventing any information leakage.

Various QNN ans\"{a}tze can be designed in search of a solution to some quantum machine learning task.  In general, all QNNs are formed by a series of parameterized quantum gates acting on some register of quantum and optionally classical bits. However, they can be broadly categorized into three `species': dissipative QNNs~\cite{bee_tra_2020}, in which information is transferred to a new register of qubits at each layer (analogous to a classical feed-forward neural network); convolutional QNNs~\cite{car_gen_2022, con_qua_2019}, where some number of qubits are measured at each layer to reduce the dimension of the data; and `standard' QNNs~\cite{kordzanganeh2023exponentially,kordzanganeh2023parallel,sagingalieva2023hybrid,sedykh2024hybrid,anoshin2024hybrid}, in which the number of qubits remains constant throughout. Of the three, the latter type of QNN is the most commonly used throughout the literature for quantum machine learning tasks. For that reason, a similar architecture is chosen for the present study. Our ansatz, shown in Fig.~\ref{fig:qnn_ansatz}, consists of four qubits, initially prepared in the $\ket{0}$ state. Four data features, obtained after PCA, are then encoded into the angles of $R_x$ qubit rotations in the encoding layer. The feature encoding layer is followed by a layer of parametrized general rotation gates $R$ and a layer of $R_{xx}$ Ising gates between nearest-neighbors, referred to as the trainable layer. The alternating feature encoding and trainable layers are repeated $L$ times. In our ansatz, we chose to measure the output as a Pauli-Z measurement on the first qubit after an additional general parametrized rotation on this qubit. The measured value is further multiplied by a trainable prefactor, and a trainable bias is added, resulting in the final prediction of the model. By adding this minimal linear layer, we add extra flexibility to the QNN but still focus on the quantum part of the model in our benchmarks.

To assess the impact of quantum decoherence on the training process of a QNN, the strength of the three quantum noise sources described in Sec. \ref{sec:decoherence} is varied independently for several values of $p$ according to the following distribution

\begin{equation}
    \gamma = 10^{p} \qquad p \in [-3, 0],\label{eq:gamma_distribution}
\end{equation}
and the results are averaged over 20 random trajectories.

The quantum model is trained using the Lion algorithm, with a batch size of 50 for the diabetes dataset and 100 for the concrete compressive strength dataset, and the training dataset is randomly permuted at each training epoch in order to randomize the samples in each batch across epochs. The trainable parameters of the model are optimized by gradient descent using an Adam optimizer to minimize the mean squared error (MSE) loss. The Lion optimizer is initialized with a learning rate $\gamma = 0.001$, beta coefficients $\beta_1 = 0.9$ and $\beta_2 = 0.99$, and a weight decay parameter $\lambda = 0$. At each epoch, the prediction accuracy of the model is also measured by executing the model in the feed-forward mode and calculating the average MSE across the validation dataset.

\section{Results}
\label{sec:results}

The results of the training process of the model with $L$ = 5 layers, described in Sec. \ref{sec:training}, for the diabetes dataset are shown in Fig.~\ref{fig:epochal_loss}, which depicts the mean training and validation set losses at each epoch for varying strengths of the three noise channels. 

Initially, we note that very high levels of noise reduce the ability of the model to accurately characterize a dataset, which is evidenced by the reduction in the gradient of the loss curves for values of $\gamma > 0.1$. This is a consequence of the fact that the measurement outcome of the quantum circuit becomes increasingly uncorrelated with the input data features as noise in the quantum circuit increases. For example, in the case of the amplitude-damping channel, described by Eqs. (\ref{eq:ad_channel_e0}) and (\ref{eq:ad_channel_e1}), in the limit of $\gamma \to 1$, the state $\ket{0}$ will be measured with unit probability, corresponding to the value $\langle z\rangle = -1$ of the Pauli-Z observable. The depolarization channel in turn drives each qubit into an equal mixture of states $\ket{0}$ and $\ket{1}$, i.e., $1/2(\ket{0}\bra{0}+\ket{1}\bra{1})$, resulting in the output of the Pauli-Z observable equal to 0. Finally, the phase-damping noise at $\gamma \to 1$ erases quantum coherence from the qubits, leaving them in a classical mixture. This noise converts a quantum computer into a classical probabilistic device, which, for long enough circuits, again tends to output qubits in a completely mixed state.

\begin{figure*}
    \centering
    \includegraphics{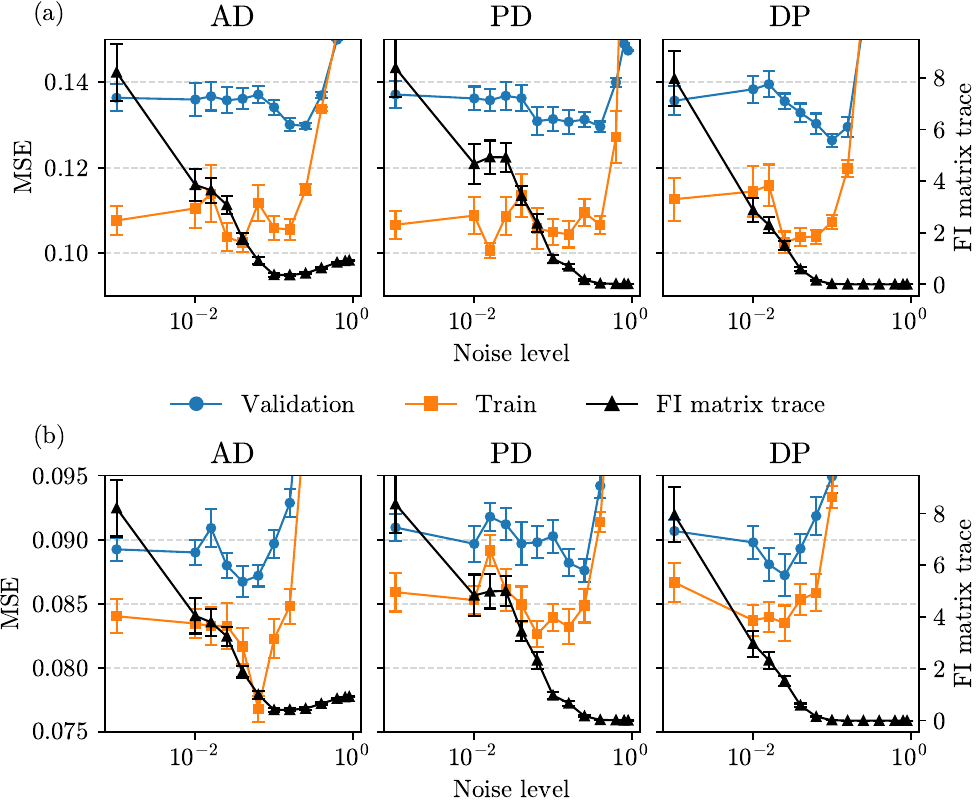}
    \caption{ Average value of the minimum validation loss obtained during training, averaged over 20 trajectories, together with the training loss at the same epochs, plotted versus the noise rate $\gamma$ for (a) the diabetes dataset and (b) the concrete compressive strength dataset. Three types of noise are considered, as indicated by the subplot titles: Amplitude-damping (AD), Phase-damping (PD), and Depolarizing (DP). The black line with triangular markers shows the trace of the Fisher Information (FI) matrix averaged over 100 random weights. The error bars indicate the standard error of the mean across trajectories. The minimum values of the validation loss are found at a non-vanishing value of the noise level.  }
    \label{fig:final_loss}
\end{figure*}

What is interesting is the behavior observed for intermediate levels of noise. Figure~\ref{fig:final_loss} demonstrates minimal validation loss versus the noise strength averaged over 20 trajectories for the diabetes dataset in (a) and the concrete compressive strength dataset in (b). The figure shows that the validation loss for both datasets and for each type of noise exhibits a minimum for some finite value of $\gamma$ greater than zero. Intuitively, one might expect that any level of noise results in a degradation in the performance of a QNN. However, the results presented here demonstrate that adding noise to a QNN can enhance its performance. Therefore, the noise level can be treated as a hyperparameter, and the optimal level can be determined using standard hyperparameter optimization methods during training.

We argue that the improved efficiency is due to improved generalization and effectively resembles the effect of regularization in "classical" machine learning. In our experiments, the quantum neural network is trained with MSE loss. Minimizing MSE is equivalent to assuming a homoscedastic Gaussian likelihood with fixed variance $\sigma^2$ for the target data $y$,
\begin{equation}
  p(y|x;\theta) \;\propto\;
  \exp\!\Big[-\tfrac{1}{2\sigma^2}\big(y-\hat y_\theta(x)\big)^2\Big],
\end{equation}
so that minimizing the negative log-likelihood reduces to
\begin{equation}
  -\log p(y|x;\theta) \;\sim\; \tfrac{1}{2\sigma^2}\,\big(y - \hat y_\theta(x)\big)^2 .
\end{equation}
In this setting, the generalized Gauss–Newton matrix
\begin{equation}
  F(\theta) \;=\; \frac{1}{\sigma^2}\,
  \mathbb{E}_x\!\left[\nabla_\theta \hat y_\theta(x)\,
                      \nabla_\theta \hat y_\theta(x)^\top\right],
\end{equation}
averaged over a set of random points $\{x\}$, coincides with the empirical Fisher information (the expected outer product of gradients) and approximates the Hessian of the empirical risk near a good fit, where residuals are small. Thus, the Fisher spectrum is a valid measure of the curvature of the MSE loss surface, exactly as in the negative log-likelihood case~\cite{Amari1998Natural,Martens2014Deep,Grosse2016Scaling}.

The trace of the Fisher equals the sum of its eigenvalues and quantifies the model’s total sensitivity of predictions to infinitesimal parameter perturbations. Large traces and large top eigenvalues correspond to sharp minima that are prone to overfitting, whereas smaller values indicate flatter basins that are typically more robust and generalise better~\cite{Hochreiter1997Flat,Dinh2017SharpMinima}.

We therefore augment the performance plots in Fig.~\ref{fig:final_loss} with the Fisher trace averaged over 100 random initializations of the QNN parameters. When computing the Fisher matrices, we exclude the final bias parameter of the linear layer, since $\partial \hat y/\partial b = 1$ deterministically contributes one to the trace. Figure~\ref{fig:final_loss} shows that, as the circuit noise level~$\gamma$ increases, the Fisher trace decreases for each type of noise. This behaviour provides a mechanistic signature of regularisation under MSE training: injected decoherence suppresses overly sensitive directions in parameter space, compresses the Fisher spectrum, and steers training towards flatter solutions that coincide with improved validation loss.

We also observe that the training loss, evaluated at the same epochs where the minimum validation loss is reached in Fig.\ref{fig:final_loss}, decreases as the noise level increases. The location of this minimum closely coincides with the minimum of the validation loss. Importantly, this effect cannot be attributed to improved trainability of the training objective: as shown in Fig.\ref{fig:epochal_loss}, the overall training loss actually increases at the noise levels where the minimum validation loss is considerably improved. Instead, we attribute this behavior to enhanced generalization. Noise suppresses sharp directions in the loss landscape, as evidenced by the monotonic decay of the Fisher trace in Fig.~\ref{fig:final_loss}, which delays the onset of overfitting. As a result, the model continues to generalize well for longer training times, as can also be seen in Fig.\ref{fig:epochal_loss}, leading to a lower validation loss and a correspondingly smaller training loss at the epoch where the validation minimum is achieved.

\section{Quantum Noise Regularization}

In the context of the results outlined in Fig.~\ref{fig:final_loss}, we propose that noise can be considered as a type of `quantum regularization.' In other words, the noise described in Sec. \ref{sec:decoherence} has the ability to reduce the degree of overfitting and improve the model's ability to predict unseen data. In order to leverage this effect, we propose a method that allows for the level of noise in a QNN to be treated as a hyperparameter of the optimizer applied in the training phase by augmenting the quantum circuit with additional quantum gates or by the introduction of artificial noise. This method is applicable to both quantum emulators and quantum processing units:

\begin{enumerate}
    \item An initial hybrid or fully quantum ansatz is specified based on some problem-specific criteria.
    \item Sources of quantum noise describing one or more quantum decoherence channels are then inserted into the ansatz after each gate operation.
    \item An input parameter is provided to these sources, enabling the adjustment of the level of decoherence $\gamma$.
    \item The value of the input parameter is then chosen based on performance metrics of the QNN applied to the problem, e.g
    \begin{itemize}
        \item Through the use of some optimization routine that minimizes the validation loss or some equivalent metric.
        \item Through a systematic selection of the noise parameter based on holistic constraints.
    \end{itemize}
\end{enumerate}

In essence, the method seeks to adapt a variational quantum algorithm in such a way that the degree of noise in the circuit becomes a hyper-parameter of the ansatz.

\subsection{Introducing Additional Noise}

In the procedure outlined above, the method by which noise is introduced to a quantum circuit was defined abstractly. In this section, we seek to outline some concrete ways in which this can be achieved.

In the case of quantum emulators, a specific level of noise can be achieved straightforwardly within the density matrix representation using the Kraus operator formalism for a given quantum channel. Some examples of these quantum channels were given in Sec. \ref{sec:decoherence}. In this case, the boxes depicted in Fig.~\ref{fig:control_architecture} take the form of parameterized Kraus operators.  

On the other hand, controlling the amount of noise above some minimum (set by the intrinsic level of hardware noise) for a physical QPU is not straightforward. When comparing the optimum levels of noise across different circuit depths to the measured coherence times for popular quantum processing units (Table \ref{tab:qpu_noise}), it is apparent that the level of noise present on many quantum devices is lower than the level of noise required to achieve optimum performance for some quantum machine learning problems. These values are shown, respectively, in Tables \ref{tab:optimum_noise} and \ref{tab:qpu_noise} for each of the three channels. Moreover, the optimum level of noise for producing the regularization phenomenon is observed to be approximately constant with the increasing number of ansatz layers.

\begin{table*}
    \centering
    \begin{tabular}{|c|c|c|c|c|c|c|c|}
        \hline
        Noise Channel & Mean Optimum Noise & $L=3$ & $L=4$ & $L=5$ & $L=6$ & $L=8$ & $L=10$ \\
        \hhline{|=|=|=|=|=|=|=|=|}
        $\gamma_\text{AD}$ & \num{1.00E-02} & \num{1.78E-02} & \num{1.78E-02} & \num{1.78E-02} & \num{1.00E-02} & \num{1.00E-02} & \num{1.31e-2} \\
        $\gamma_\text{PD}$ & \num{3.16E-02} & \num{3.16E-02} & \num{5.62E-02} & \num{3.16E-02} & \num{3.16E-02} & \num{1.78E-02} & \num{3.38e-2} \\
        $\gamma_\text{DP}$ & \num{5.62E-03} & \num{5.62E-03} & \num{1.00E-02} & \num{5.62E-03} & \num{5.62E-03} & \num{5.62E-03} & \num{6.50e-3} \\
        \hline
    \end{tabular}
    \caption{The average optimum noise parameter for the three noise channels across circuits of variable depth with $3 \leq L \leq 10$ layers.}
    \label{tab:optimum_noise}
\end{table*}

\begin{table*}
    \centering
    \renewcommand{\arraystretch}{1.2}
    \begin{tabular}{|c|c|c|c|c|c|}
        \hline
        Hardware & $T_1$ & $T_2$ & $T_G$ (2Q) & $\gamma_\text{AD}$ & $\gamma_\text{PD}$ \\
        \hhline{|=|=|=|=|=|=|}
        Rigetti Aspen M-3~\cite{rig_rig_2022}       & \SI{25.0}{\micro\second} & \SI{28.0}{\micro\second} & \SI{240}{\nano\second}  & \num{9.55e-3} & \num{8.54e-3} \\
        IonQ Aria~\cite{ion_ama_2023, ion_ion_2023} & \SI{10}{\second}         & \SI{1}{\second}          & \SI{600}{\micro\second} & \num{6.00e-5} & \num{6.00e-4} \\
        Google Sycamore~\cite{aru_qua_2019}         & \SI{15}{\micro\second}   & \SI{19}{\micro\second}   & \SI{12}{\nano\second}   & \num{8.00e-4} & \num{6.31e-4} \\
        IBM Osprey Seattle~\cite{ibm_ibm_2023}      & \SI{85.9}{\micro\second} & \SI{63.0}{\micro\second} & \SI{635}{\nano\second}  & \num{7.37e-3} & \num{1.00e-2} \\
        \hline
    \end{tabular}
    \caption{The average amplitude damping and phase damping noise for various QPUs, obtained via Eqs. (\ref{eq:gamma_ad}) and (\ref{eq:gamma_pd}) from coherence times provided by the vendors.}
    \label{tab:qpu_noise}
\end{table*}

Decoherence rates can be obtained experimentally and thus are commonly used by QPU vendors as a metric for noise robustness. Two decoherence rates are usually quoted, the so-called relaxation time $T_1$ and the dephasing time $T_2$, which are closely related to two of the fundamental channels outlined in Sec. \ref{sec:decoherence}. Given the single qubit gate execution time, $T_G$, and the value of $T_1$ or $T_2$, one can estimate the value of the noise parameter $\gamma$ for the \textit{amplitude-damping} and \textit{phase-damping} channels respectively. In practice, this means the degree of amplitude damping and phase damping noise can be increased by artificially extending the gate execution time through the addition of idle time following each gate operation. 

In the case of the \textit{depolarizing} channel, additional noise can be introduced by mimicking the effect of the depolarizing channel with single-qubit Pauli operators. The Kraus operators in Eqs. (\ref{eq:dp_channel_e0}) - (\ref{eq:dp_channel_e3}) correspond to the identity operator and the three Pauli operators (\I, \X, \Y, \Z). Thus, the effect of additional depolarizing noise can be achieved using a random number generator that applies one of the Pauli operators with probability $\gamma/3$ or the identity operator with probability $1 - \gamma$.

\begin{align}
    \gamma_\textrm{AD} &= 1 - \exp \left( -\frac{T_G}{T_1} \right) \label{eq:gamma_ad} \\
    \gamma_\textrm{PD} &= 1 - \exp \left( -\frac{T_G}{T_2} \right) \label{eq:gamma_pd}
\end{align}

\section{Numerical simulation of using quantum noise regularization for QNN implemented on IBM quantum computer}

\begin{figure*}
    \centering
    \includegraphics{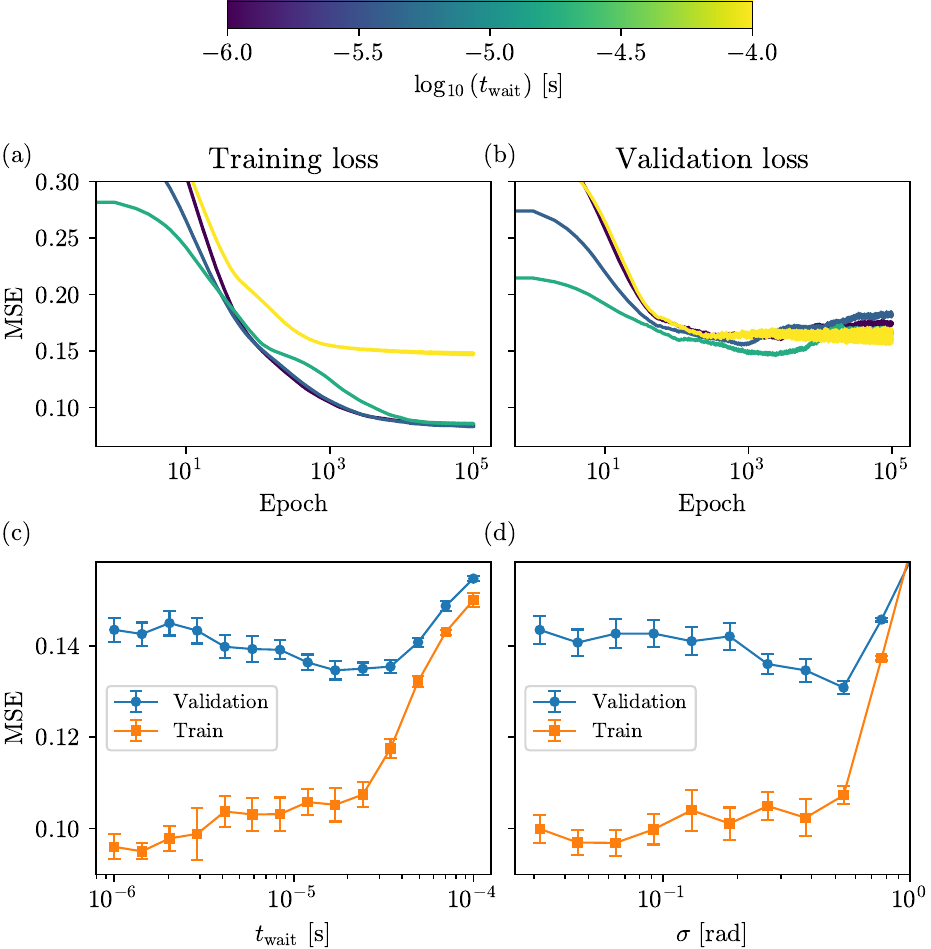}
    \caption{Training (a) and validation (b) losses from noisy simulations of a QNN on the IBM \texttt{kingston} processor using the diabetes dataset. Idle waiting times $t_\mathrm{wait}$ are varied as indicated by line color. Panels (c) and (d) show the average minimum validation loss and the corresponding training loss versus $t_\mathrm{wait}$ and the strength of stochastic $x$–rotations $\sigma$, respectively. Results are averaged over 20 trajectories trained for $10^5$ epochs; error bars indicate the standard error of the mean. Both approaches yield nonzero optimal values of $t_\mathrm{wait}$ and $\sigma$, demonstrating the potential of noise-induced regularization on realistic hardware.}
    \label{fig:IBM_regularisation}
\end{figure*}


To assess whether the proposed noise-induced regularisation can be
executed on current or near-future hardware, we performed a noisy simulation
of our quantum neural network (QNN) tailored to the
$156$-qubit \texttt{ibm\_kingston} processor (Heron r2 architecture).

The employed  circuit keeps the number of qubits and the layer pattern of the circuit presented in Fig.~\ref{fig:qnn_ansatz}, but every gate
is replaced by Kingston’s native gates as follows:
\begin{align}
  R(\theta_1,\theta_2,\theta_3)
    &\longrightarrow R_{z}(\theta_{1})\,R_{x}(\theta_{2})\,R_{z}(\theta_{3}),\label{eq:nativeRxz}\\
  R_{xx}(\theta)        &\longrightarrow R_{zz}(\theta).
\end{align}

We derive all noise parameters from the latest public calibration snapshot
(July~2025) and use the \emph{medians} to obtain a representative ``typical''
device.  Table~\ref{tab:kingston_medians} summarises the values and the way
they enter the simulator.

\begin{table}[ht]
    \centering
    \caption{Median calibration values for \texttt{ibm\_kingston} and how they are mapped to noise channels.}
    \label{tab:kingston_medians}
    \begin{tabular}{|c|c|c|c|}
        \hline
        Quantity & Median value & Applied to & Simulator channel(s) \\
        \hhline{|=|=|=|=|}
        1-qubit EPC ($R_x$)   & $2.335\times10^{-4}$ & every physical 1-q gate & depolarising (prob.\ $p=\text{EPC}$) \\
        2-qubit EPC ($R_{zz}$)& $2.137\times10^{-3}$ & each entangler          & two-qubit depolarising \\
        Gate durations        & $35\,$ns ($R_x$), $250\,$ns ($R_{zz}$) & relaxation time of each gate & amplitude + phase damping \\
        Coherence times       & $T_1=224.4\,\mu$s,\; $T_2=120.5\,\mu$s & all qubits & parameters of relaxation channels \\
        Virtual $R_z$         & error-free & — & frame shift only \\
        \hline
    \end{tabular}
\end{table}

Every physical gate is thus followed by
\emph{(i)}~a depolarising channel with the appropriate error-per-Clifford (EPC)
and \emph{(ii)}~a thermal-relaxation channel whose Kraus parameters are computed
from the gate duration and the median $T_1$ and $T_2$.


We considered two approaches to introduce noise into the circuit in order to emulate the regularisation protocol. In the first approach, we insert an additional idle period of length~$t_{\mathrm{wait}}$ after each gate on qubits on which this gate was applied, as indicated in Fig.~\ref{fig:control_architecture}.
During this pause, the qubits undergo the amplitude- and phase-damping
processes according to the values of $T_1$ and $T_2$.

In the second approach, we append after every physical gate an additional stochastic $R_x(\delta\phi)$ gate on each qubit that participated in that gate, where the angle is drawn independently as $\delta\phi\sim\mathcal{N}(0,\sigma^2)$. Averaging over the Gaussian distribution produces an effective dephasing in the eigenbasis of the generator of $R_x$, i.e., the Pauli-$X$ operator. Equivalently, in a Lindblad description, this corresponds to a single jump operator proportional to the gate’s Hamiltonian, $X$ in our case, with an induced dephasing probability $p=(1-e^{-\sigma^2/2})/2$. In our hardware-realistic simulations, we also account for the native errors of applying such an $X$ gate, similar to the circuit gates. This approach can equally use any other parametrized gates available on a considered hardware.

Figures~\ref{fig:IBM_regularisation}(a)–(b) show the averaged training and validation loss trajectories for the idle–wait approach on the diabetes dataset. Consistent with the modeled-noise results in Fig.~\ref{fig:epochal_loss}, the validation loss attains lower values at intermediate idle times $t_{\mathrm{wait}}$. Further, Figs.~\ref{fig:IBM_regularisation}(c)–(d) give, for each noise level, the minimum validation loss per trajectory (averaged over runs) together with the training loss at the same epoch, as functions of $t_{\mathrm{wait}}$ in panel~(c) and the rotation-noise standard deviation $\sigma$ in panel~(d). Clear optima emerge around $t_{\mathrm{wait}}\!\approx\!1$–$2\times10^{-5}\,\mathrm{s}$ and $\sigma\!\approx\!0.6$. These results demonstrate that using the noise as a hyperparameter can improve validation loss of quantum models at noise levels compatible with today’s superconducting hardware.

\section{Conclusion}

This work presents a study of the effect of the three most common noise channels on QNNs. We explore how adjusting the levels of amplitude damping, phase damping, and depolarizing noise within a quantum circuit influences a QNN's capacity to learn and generalize data. For example, we considered regression tasks based on the diabetes and concrete compressive strength datasets, where, by tuning the noise level in the circuit, we obtained an improvement of the mean squared error loss. Our observation reveals that introducing a certain degree of decoherence throughout the circuit can effectively prevent overfitting. Building upon these findings, we propose a technique that leverages noise to enhance the generalization performance of QNNs. Finally, we numerically demonstrated the effectiveness of our approach by numerically simulating noise-induced regularization of QNN implemented on the IBM Kingston quantum computer using the corresponding noise model.

The practical implementation of our technique presents challenges, particularly when transitioning from optimization on simulation to quantum hardware. While we have shown for a particular task that the optimal noise level required to prevent overfitting is lower than the noise level of the hardware, there may be instances where the opposite holds true. In such cases, introducing additional noise would degrade performance rather than enhance it. However, this issue might be solved by using a different quantum circuit architecture or by adjusting the level of error mitigation. These findings suggest that it would be valuable to explore the application of this noise-leveraging technique within hybrid quantum neural networks, particularly for advancing performance in industrial tasks such as image processing~\cite{sagingalieva2023hyperparameter,lusnig2024hybrid,senokosov2024quantum}, time series forecasting~\cite{sagingalieva2025photovoltaic,kurkin2025forecasting}, and planning~\cite{rainjonneau2023quantum,haboury2023supervised}.

\bibliography{main}
\bibliographystyle{unsrt}

\end{document}